# Novel Heusler Materials for Spintronic Applications:
# Growth, Characterizations and Applications

Ravinder Kumar[1] and Sachin Gupta[1,2*]

[1]Department of Physics, Bennett University, Greater Noida 201310, India

[2]Center of Excellence in Nanosensors and Nanomedicine, Bennett University, Greater Noida 201310, India

*Corresponding author email: gsachin55@gmail.com

## Abstract

Spintronics is a rapidly evolving technology that utilizes the spin of electrons along with their charge to enable high-speed, low-power and non-volatile electronic devices. The development of novel materials with tailored magnetic and electronic properties is critical to exploit the full potential of spintronic applications. Among these, Heusler alloys stand out due to their tunable multifunctional properties. This review presents a comprehensive overview of various Heusler-based materials—including half-metallic ferromagnets, spin gapless semiconductors, magnetic semiconductors, spin semimetals, and nearly zero-moment materials—focusing on their synthesis, structural and magnetic characterizations, and transport behavior. The role of crystal structure, and structural disorder in governing their magnetic and electronic properties is discussed in detail. Emphasis is placed on experimental results and their implications for spintronic devices. By bringing together recent advancements, the review highlights the critical role of Heusler alloys in advancing the next-generation spintronic technologies and outlines future directions for their integration in practical applications.

## 1. Introduction

Spintronics is a rapidly growing field, which harnesses both the spin and electric charge of electron, holds immense promise for revolutionizing electronic devices.[1–3] Unlike traditional electronics, which rely solely on charge, spintronic devices exploit the electron's spin property to achieve various advantages such as non-volatility, high storage density, low power consumption, and faster operation.[4–6] The field gained significant attention with the discovery of giant magnetoresistance (GMR) by Albert Fert and Peter Grünberg, who demonstrated a significant resistance change in multilayer magnetic systems under a magnetic field.[7,8] Fig. 1 shows a multilayer system consisting of ferromagnetic iron (Fe) layers separated by very thin non-magnetic Chromium (Cr) layers. Awarded the 2007 Nobel Prize in Physics, this breakthrough paved the way for compact, efficient hard disk drives. Building on this foundation, the subsequent development of spin valves provided a practical means to harness GMR for sensing and data-reading technologies, further expanding its real-world applications.

The evolution of spintronics continued with the introduction of magnetic tunnel junctions (MTJs), which utilize tunneling magnetoresistance (TMR) to achieve greater efficiency and scalability.[9–11] By sandwiching a thin insulating layer (e.g. $Al_2O_3$ and MgO) between two magnetic layers, MTJs allow precise control of spin currents, laying the groundwork for advanced applications such as magnetic random-access memory (MRAM).[12] The potential of this technology became tangible in 1999 when IBM unveiled the first MRAM prototype, demonstrating faster data access, enhanced durability, and lower power consumption compared to conventional DRAM and SRAM.[13,14] This marked a pivotal moment, showcasing spintronics as a viable contender for future memory solutions.

Further progress came with the advent of spin-transfer torque (STT), a mechanism that manipulates magnetic states using spin-polarized currents instead of external magnetic fields. This breakthrough has fueled the rise of STT-MRAM, offering improved energy efficiency and scalability for modern memory systems. More recently, the exploration of Heusler alloys, topological insulators and two-dimensional materials like graphene has opened exciting new avenues. These materials enable spin currents to propagate over longer distances with minimal energy loss, promising ultra-low-power spintronic devices and potential integration with quantum computing (studies on 2D materials and topological insulators). Additionally, phenomena such as the spin Hall effect and spin-orbit coupling have enriched the field, forging connections with quantum technologies and neuromorphic computing.

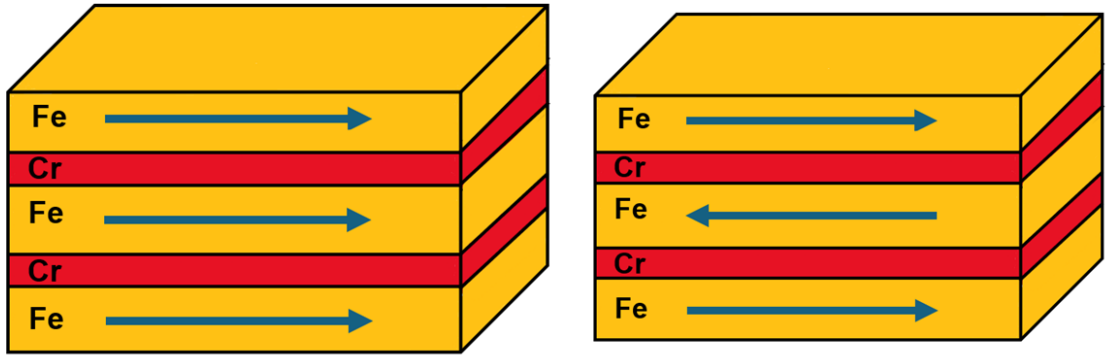

**Fig. 1.** Schematic of multilayer systems consisting of alternating magnetic and non-magnetic layers.[5]

As a dynamic and interdisciplinary domain, spintronics merges quantum mechanics, materials science, and electronics to address contemporary technological challenges. Its ongoing evolution, driven by the pursuit of advanced materials and novel physical phenomena, continues to push the boundaries of what's possible, positioning it as a cornerstone for the future of computing and beyond.

## 2. Spintronic materials

The progress of any emerging technology relies on the theoretical prediction of novel materials and their experimental realization. Various material families—such as topological,[15–19] semiconductors,[20–22] half-metals,[23] two-dimensional materials,[24–27] and fully compensated ferrimagnets[28,29] have been theoretically predicted and experimentally investigated for their promising properties in spintronic applications. In this review, our focus will be on Heusler alloys, which have emerged as potential candidates for spintronic devices due to their multifunctional properties.

### 2.1 Heusler alloys, their types and crystal structures

The discovery of Heusler alloys goes back to 1903 when F. Heusler,[30] a German chemist, identified a very interesting intermetallic ferromagnetic compound ($Cu_2MnAl$) made up of copper (Cu), manganese (Mn), and aluminum (Al). Subsequently, more Cu- and Mn-based ferromagnetic materials were discovered in combination with other elements such as antimony (Sb), bismuth (Bi), and tin (Sn). Although these individual elements are non-magnetic, the resulting alloys exhibited ferromagnetic properties. This finding contradicted the earlier belief that magnetism is limited to only a few elements like iron, cobalt, and nickel. In contrast, Heusler's discoveries demonstrated that non-magnetic elements could exhibit magnetism when arranged in specific structures and interactions within intermetallic compounds. This insight has inspired the search for new materials with tunable magnetic properties.

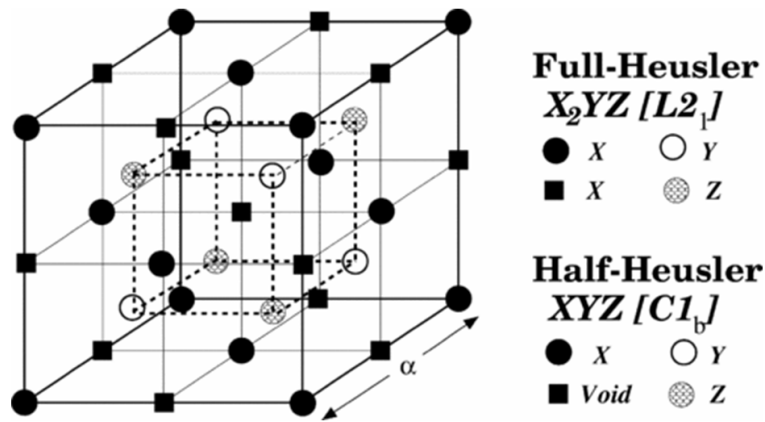

**Fig. 2**. Schematic for full Heusler alloy and half Heusler alloy lattice structures. The full Heusler alloy has four interpenetrating fcc sublattices whereas only three of them are occupied in half Heusler alloy. Reproduced with permission from E. Şaşıoğlu *et al.*, Phys. Rev. B 72 (2005) 184415, doi:10.1103/PhysRevB.72.184415. © 2005 American Physical Society.[31]

Heusler alloys can be classified into two primary categories based on their stoichiometric composition; half Heusler (1:1:1), and full Heusler (2:1:1) alloys. Full Heusler alloys can be further sub-divided in inverse and quaternary Heusler alloy. Fig. 2 depicts four interpenetrating face centered cubic (fcc) sublattices and their occupancy for full and half Heusler alloys.[31]

Half Heusler alloys are represented by a general formula *XYZ* (where *X* and *Y* are transition metal elements and *Z* is a main group element) and have a non-centrosymmetric $C1_b$ (space group $F\bar{4}3m$, *no.* 216) type Cubic structure, e.g. NiMnSb, CoMnSb and PtMnSb.[32] In $C1_b$ structure, out of four, only three fcc sub-lattices are occupied by $X$, $Y$, and $Z$ atoms with respect to Wyckoff positions 4a (0, 0, 0), 4b (1/2, 1/2, 1/2), and 4c (1/4, 1/4, 1/4).[33] Full Heusler alloys are represented by the general formula $X_2YZ$ and have the $L2_1$ structure (space group $Fm\bar{3}m$ no. 225) with $Cu_2MnAl$ serving as a prototype. In $L2_1$ structure, all four interpenetrating fcc sublattices are occupied as shown in Fig. 2.[34] $X$, $Y$, and $Z$ atoms in this structure occupy Wyckoff positions, 8c (1/4, 1/4, 1/4), 4a (0, 0, 0) and 4b (1/2,1/2, 1/2), respectively.[33] Examples of full Heusler alloys are $Co_2Mnsb$, $Co_2TiSn$, $Co_2VAl$ etc.[35,36]. Inverse Heusler alloys can be represented by $X_2YZ$, however in these materials the atomic number of $Y$ atoms is higher than the one of the $X$ atoms.[37] In this configuration $X$ is more electropositive than $Y$. This structure also has four interpenetrating fcc sublattices. The prototype structure for these materials is $CuHg_2Ti$ [37] which crystallizes in the space group $F\bar{4}3m$ (no. 216) for example $Sc_2MnAl$, $Ti_2VAl$ etc.[38] When both the $X$ atoms in $X_2YZ$ are different transition metal elements, it results in a new Heusler alloy, known as quaternary Heusler alloy, represented by a general formula, $XX'YZ$ (Fig. 3). In these alloys, LiMgPdSn is considered as a prototype structure with space group $F\bar{4}3m$ (no. 216) and called as a $Y$-type structure. Examples of quaternary Heusler alloys are CrVTiAl, CoFeCrGe and CoMnCrAl etc.[39,40] If there is an exchange in the lattice sites of $X$ and $X'$, this leads to $L2_1$ type structure. Three distinct types of structural compositions for the $Y$-type structure are possible based on the position of the different lattice sites named as $Y_I$, $Y_{II}$, and $Y_{III}$.[41] If $Z$ atom is fixed at 4a (0,0,0) Wyckoff position, then the three possible energetically non-degenerate configurations are as following: [33,39]

$Y_I \rightarrow X$ at 4d (3/4,3/4,3/4), $X'$ at 4c (1/4,1/4,1/4) and $Y$ at 4b (1/2,1/2,1/2) b).

$Y_{II} \rightarrow X$ at 4b (1/2,1/2,1/2), $X'$ at 4c (1/4,1/4,1/4) and $Y$ at 4d (3/4,3/4,3/4) c).

$Y_{III} \rightarrow X$ at 4d (3/4,3/4,3/4), $X'$ at 4b (1/2,1/2,1/2) and $Y$ at 4c (1/4,1/4,1/4).

## 2.2 Disorder in Heusler alloys

The arrangement of atoms within the unit cell plays a crucial role in Heusler alloys, as it significantly influences their magnetic, electrical, and related properties, which in turn can impact device performance.[35] Therefore, investigating structural disorders in Heusler alloys is an essential step. It has been observed that when all atoms occupy their designated positions in the unit cell, the structure is considered ordered. However, in many cases, atoms may swap positions with one another, leading to disorder in the Heusler alloy structure. Depending on the nature of the atomic swapping, different types of disorder in Heusler alloys are described below, illustrated in Fig. 4, and summarized in Table I.[33,42,43]

(a) *B*2 disorder: when *Y* and *Z* atoms swap their positions in the unit cell.

(b) *DO₃* disorder: when *X* and *Y* atoms swap their positions in the unit cell.

(c) *A*2 disorder: when *X*, *Y*, and *Z* atoms are mix and occupy random positions.

(d) *B*32a disorder: when *X*-atoms of one of fcc sublattice are mix *Y*-atoms, on the other hand second sublattice *X*- atoms are mixed with *Z*-atoms.

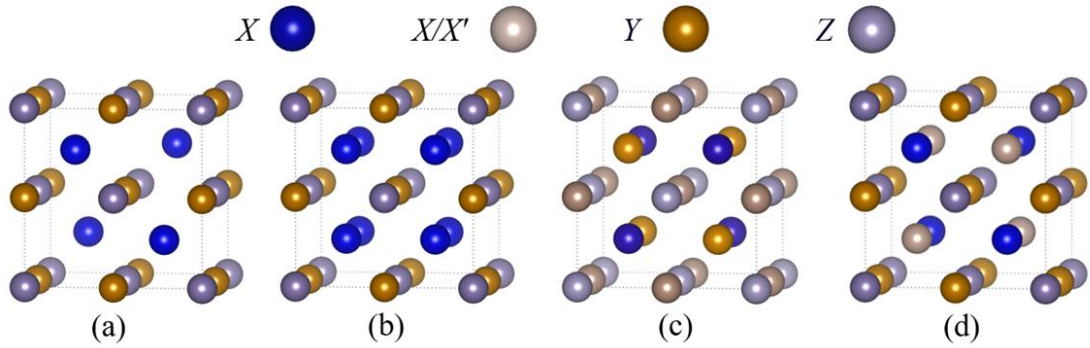

**Fig. 3.** Unit cell structure representation for (a) Half Heusler, (b) Full Heusler, (c) Inverse Heusler, and (d) quaternary Heusler alloys.[33]

**Table I**: Prototype structure, general formula, site occupancy, structure type and the space group for different types of Heusler alloys.[33]

| Prototype | General formula | Site occupancy | Structure | Space group |
|---|---|---|---|---|
| LiMgPdSn | *XX′YZ* | *X, X′, Y, Z* | *Y* | $F\bar{4}3m$ (no. 216) |
| Cu₂MnAl | *X₂YZ* | *X=X, Y, Z* | *L2₁* | $Fm\bar{3}m$ (no. 225) |
| CsCl | *X₂Y₂* | *X=X, Y=Z* | *B2* | $Pm\bar{3}m$ (no. 221) |
| BiF₃ | *X₃Z* | *X=X=Y, Z* | *DO₃* | $Fm\bar{3}m$ (no. 225) |
| W | *X₄* | *X=X=Y=Z* | *A2* | $Im\bar{3}m$ (no. 229) |

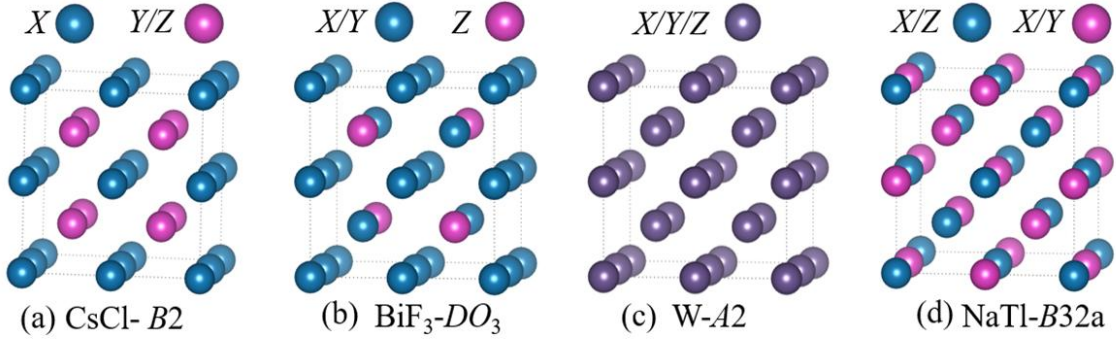

**Fig. 4.** Representation of various types of atomic disorders in Heusler alloys.[33]

Disorder in Heusler alloys can be identified using X-ray diffraction (XRD) patterns by analyzing the reflections from different crystal planes. Among these, two superlattice peaks (111) and (200) along with the principal peak (220), provide insight into the type of disorder present in the Heusler structure. The structure factors for the (111), (200), and (220) reflection peaks of quaternary Heusler alloys can be expressed as]:[40]

$$F_{111} = 4[(f_Y - f_Z) - i\,(f_X - f_{X'})] \tag{1}$$

$$F_{200} = 4[(f_Y + f_Z) - (f_X - f_{X'})] \tag{2}$$

$$F_{220} = 4[(f_Y + f_Z + f_X + f_{X'})] \tag{3}$$

$f_X$, $f_{X'}$, $f_Y$ and $f_Z$ represent the atomic scattering factors for $X$, $X'$, $Y$ and $Z$ atoms, respectively. Depending upon the swapping of atoms, structure factors can be zero, and the corresponding peak will disappear from the XRD pattern. For example, CoRuMnGe crystallizes in $L2_1$ type structure and do not fit in $Y$-type ordered structure as can be seen in inset of Fig. 5(a).[44] The XRD pattern of CoRuMnGe fits well with 50% disorder between Co and Ru sites i.e. the 4c and 4d sites are equally probable for Co and Ru atoms. Thus, due to 50% swap disorder between tetrahedral site atoms, the crystal symmetry reduces to $L2_1$. As can be seen in Fig. 5(b), the Rietveld refinement of XRD pattern for CoRuVAl shows a 50 % disorder between Co & Ru sites and V & Al sites which results in $B2$ disorder. The occupancy of 4c & 4d sites are equally probable for Co & Ru atoms and the occupancy of 4a & 4b sites are equally probable for V & Al. Complete disappearance or a low intensity of superlattice peak (111) can be observed in the XRD pattern due to $B2$ disorder in CoRuVAl.[44] Rietveld refinement of XRD pattern for CoFeRuSn shown in Fig. 5(c) clearly shows the intensity of (111) reflection is higher than the (200). Here, the exchange of lattice sites of $X$ or $X'$ with $Y$ or $X'$ with $Z$ may occur, which results in a small amount of $DO_3$ type disorder.[45] A complete $A2$ disorder in the structure of MnCrVAl can be seen in Fig. 5(d). Random distribution of $X$, $X'$, $Y$ and $Z$ atoms produce Tungsten (W) type crystal structure and therefore (111) and (200) peaks disappear in the XRD pattern. The absence of superlattice peaks (111) and (200) confirm the $A2$ disorder present in the MnCrVAl.[46]

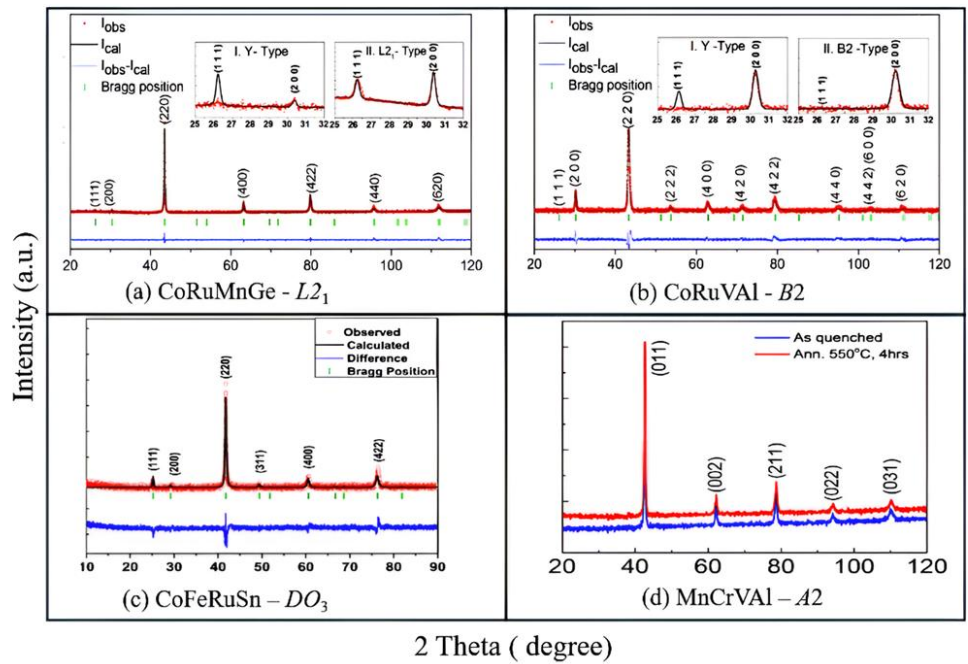

**Fig. 5.** X-ray diffraction pattern of disordered material (a) CoRuMnGe-$L2_1$, (b) CoRuVAl-$B2$, (c) CoFeRuSn-$DO_3$, (d) MnCrVAl-$A2$. Reproduced with permission from D. Rani *et al.*, J. Magn. Magn. Mater. 492 (2019) 165662, doi.org/10.1016/j.jmmm.2019.165662. © 2019 Elsevier [44], R. Kumar and S. Gupta, Appl. Phys. Lett. 124 (2024) 022402, doi.org/10.1063/5.0178437. © 2024 AIP publishing [45], P. Kharel *et al.*, AIP Adv. 7 (2017) 056402, doi.org/10.1063/1.4972797. © AIP Publishing (Creative Commons CC BY).[46]

## 3. Novel material properties for spintronic applications

Heusler alloys are well known for their multifunctional properties.[47–50] These materials exhibit tunable electronic and magnetic behaviors, enabling the design of new compounds with enhanced functionality through modifications such as doping, elemental substitution, application of pressure or strain, and incorporation of spin-orbit coupling. Several Heusler alloys—particularly

Co-based ones—demonstrate high Curie temperatures, making them promising candidates for room-temperature spintronic applications. In this section, we will discuss the structural, magnetic, and transport properties of selected Heusler materials that exhibit unique characteristics, enabling them as preferred candidates for spintronic devices.

### 3.1 Half metallic materials

Half metallic materials show unique band structures, in which one spin channel (say majority, spin up) show metallic properties, while another spin channel (say minority, spin down) shows gapped state at Fermi level, indicating semiconducting/insulating nature. The density of states (DOS) for half-metallic material is shown in Fig. 6(a). due to one conducting and another semiconducting/insulating channel, these materials result in 100% spin polarization.[23] NiMnSb half Husler alloy was discovered to be half metallic by de Groot *et al.*[23] A number of Heusler materials have been predicted to show half metallicity and later realized experimentally.[36,39,45,51–54] In the following, we will discuss a few example materials.

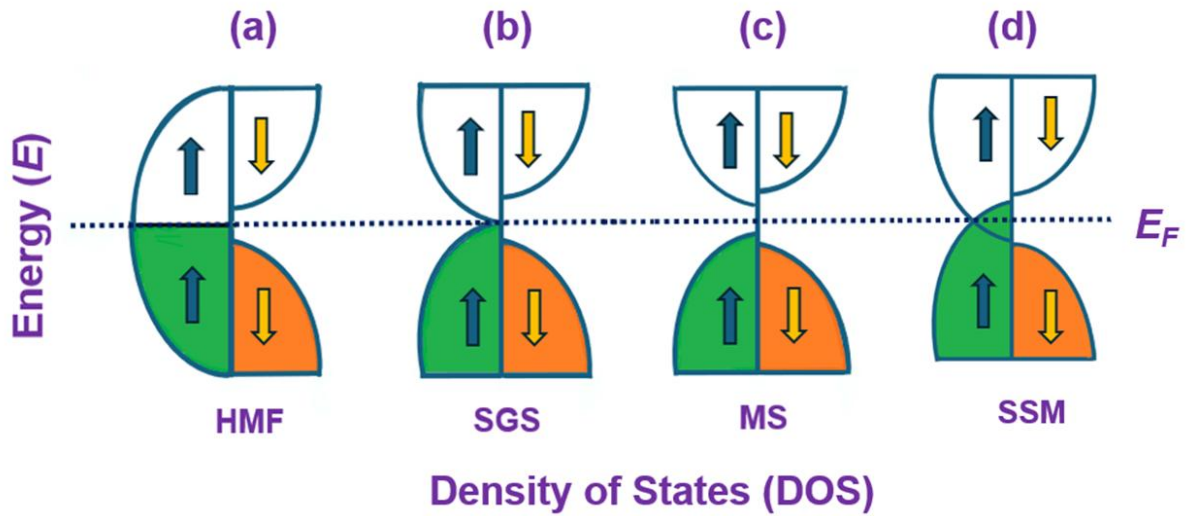

**Fig. 6.** Schematic representation of density of states (DOS) for (a) half-metallic (HMF), (b) spin gapless semiconductors (SGS), (c) magnetic semiconductors (MS), and (d) spin semi-metal (SSM).

### 3.1.1 CoFeRuSn

Recently Kumar and Gupta reported half metallic behavior in equiatomic quaternary Heusler alloy CoFeRuSn supported by magnetization and transport measurements.[45] The author synthesized quaternary Heusler alloy CoFeRuSn using arc melt technique. The structural analysis shows the material crystallizes in cubic structure with small amount of $DO_3$ disorder, as discussed in section 2.2. It can be seen from Fig. 7(a) that the material has soft ferromagnetic nature with magnetic moment 4.15 $\mu_B/f.u.$ at 4 K temperature.[45] The reported saturation magnetization deviates from the Slater-Pauling (SP) rule.[55] The deviation in the moment value from the Slater-Pauling rule has been reported in many Heusler alloys such as CoMnCrAl (Exp.-0.9 $\mu_B$; SP-1 $\mu_B$)[39], CoRuMnGe (Exp.- 4.1 $\mu_B$; SP- 4 $\mu_B$)[44], and CoRuVAl (Exp.- 0.53 $\mu_B$; SP- 1 $\mu_B$)[44], CoRuVGa(Exp.- 0.84 $\mu_B$; SP- 1 $\mu_B$)[44], $Co_2$VAl (Exp.- 1.88 $\mu_B$; SP- 2 $\mu_B$)[56], NiCoMnGe (Exp.- 4.06 $\mu_B$; SP- 6 $\mu_B$)[57]. The Curie temperature of CoFeRuSn was reported to be higher than room temperature, enabling the material for room temperature spintronic applications.

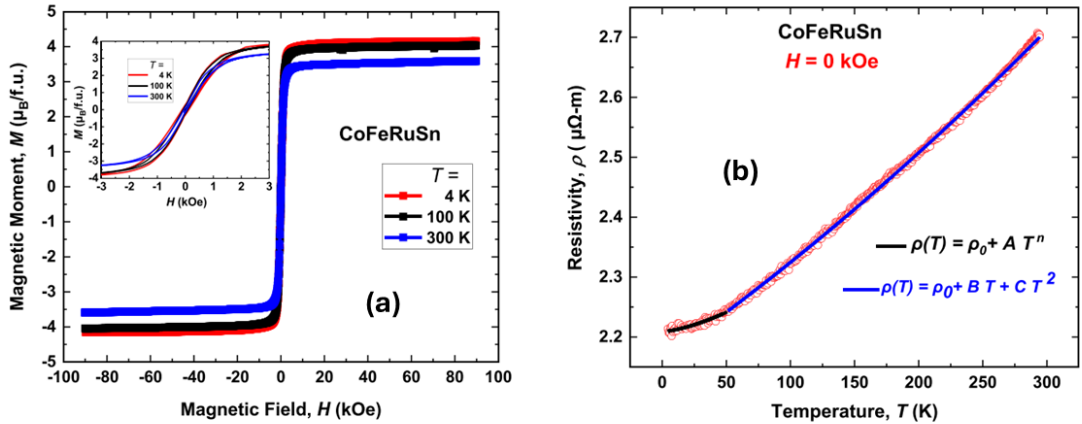

**Fig. 7.** (a) Isothermal magnetization curves (*M-H*) of CoFeRuSn at different temperatures, (b) the temperature dependence of resistivity of CoFeRuSn along with fitting. Reproduced with permission from R. Kumar and S. Gupta, Appl. Phys. Lett. 124 (2024) 022402, doi.org/10.1063/5.0178437. © 2024 AIP publishing.[45]

Transport measurements play a significant role in probing half-metallic behavior of a material indirectly. Therefore, the transport measurements as a function of temperature and field can help in identifying these materials. Temperature dependence of resistivity $\rho(T)$ in CoFeRuSn increases with temperature, indicating metallic behavior as shown in Fig. 7(b). To investigate the temperature dependence of resistivity, two different approaches have been used. In the first approach, resistivity data is fitted with the power law equation as:[58,59]

$$\rho(T) = \rho_0 + AT^n \qquad (4)$$

$\rho(T)$ represent the temperature dependence of resistivity, $\rho_0$ is the residual resistivity and $A$ is an arbitrary constant. Usually, conventional ferromagnetic materials show the quadratic temperature ($T^2$) dependence of resistivity, attributed to electron–magnon scattering.[60–62] Since the $T^2$-term is associated with single magnon scattering and is expected to be absent in half- metallic materials because of a gap in the minority spin channel.[63]

In the second approach, the resistivity due to the combined scattering contributions was considered as:

$$\rho(T) = \rho_0 + \rho_{ph} + \rho_{mag} \tag{5}$$

$\rho_{ph}$, and $\rho_{mag}$ represent resistivity due to electron–phonon scattering and due to magnonic scattering, respectively. In terms of temperature, the equation (5) can be expressed as:

$$\rho(T) = \rho_0 + BT + CT^2 \tag{6}$$

here $B$ and $C$ are arbitrary constants and magnitude of these constants can be helpful in determining the significant of the scattering.[64,65] The resistivity data was fitted in two temperature regimes. The low-temperature fitting shows a non-quadratic temperature dependence of resistivity with the value of $n$ =1.5, which indicates half-metallic nature at low temperatures.[45] The small value of the coefficient $C$ ($1.102 \times 10^{-6}$ $\mu\Omega mK^{-2}$) at higher temperatures indicates that electron–phonon scattering is more dominant than electron–magnon scattering, contributing to the degradation of half-metallicity of the sample. CoFeRuSn demonstrates half-metallicity at low temperatures however, this characteristic diminishes at higher temperatures.[45]

### 3.1.2 $Co_2Fe_{1-x}V_xGe$

Mahat *et al*. reported the half metallic type nature in $Co_2Fe_{1-x}V_xGe$ ($x = 0$ to 1, in a step of 0.125) quaternary Heusler alloy.[66] The authors synthesized $Co_2Fe_{1-x}V_xGe$ material for different values of $x$ in a step of 0.125, using arc melt technique. Stable phases were observed for $x = 0.25$ (in slow cooling) and for $x = 0.375$ (rapid ice/water mixture quenching).[66] Both samples $x = 0.25$ and 0.375 have $L2_1$ cubic structure with lattice constants 5.75Å and 5.76Å, respectively. Field dependence magnetization curves are shown in Fig. 8(a) for both the samples. Saturation magnetization in quaternary Heusler alloys follow $M = N_V - 24$ Slater-Pauling rule.[67]

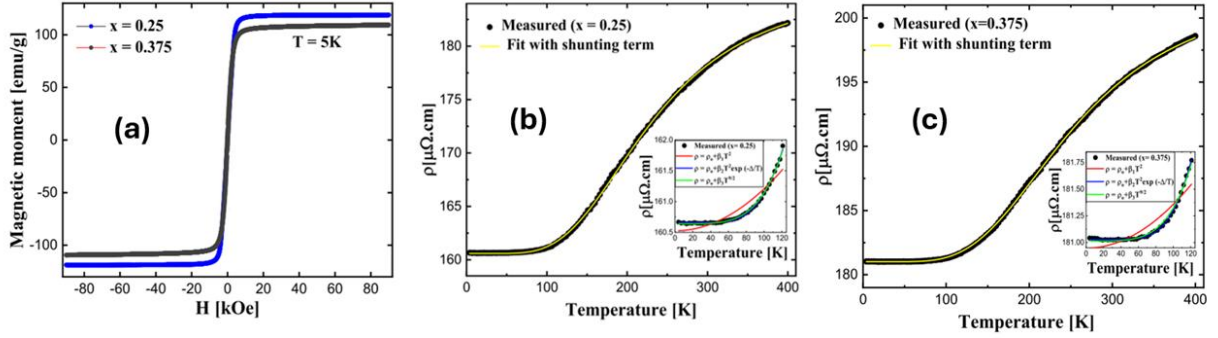

**Fig. 8.** (a) Isothermal magnetization curves of $Co_2Fe_{1-x}V_xGe$ at 5 K, (b) and (c) show the temperature dependence of resistivity for $x = 0.25$ and 0.375, respectively. Insets of (b) and (c) show fitting of resistivity data with different equations. Reproduced with permission from R. Mahat *et al.*, J. Magn. Magn. Mater. 539 (2021)168352, doi.org/10.1016/j.jmmm.2021.168352. © 2021 Elsevier.[66]

Saturation magnetic moment for $Co_2Fe_{1-x}V_xGe$ can be estimated by $M$ (x) = 6− 3x.[66] The author reported the saturation magnetic moment 5.21 $\mu_B/f.u.$ for $Co_2Fe_{0.75}V_{0.25}Ge$ and 4.78 $\mu_B/f.u.$ for $Co_2Fe_{0.625}V_{0.375}Ge$ at 5 K temperature which were in good agreement with Slater-Pauling rules. The Curie temperature ($T_c$) for $x = 0.25$ and 0.375 were found to be 820 K and 790 K, respectively, suggesting the potential of these materials for room temperature spintronic applications. The authors measured the temperature dependence of resistivity by using the Van der Pauw method[68] at zero field, shown in Fig. 8(a) and 8(b) for $x = 0.25$ and $x = 0.375$, respectively. Resistivity shows weak temperature dependency, which can be attributed to the parallel contribution of the intrinsic (temperature-dependent) resistivity and a temperature-independent limiting resistivity, commonly referred to as shunting.[69] Resistivity data of $x = 0.25$ and 0.375 were fitted by using equation (6).[70]

Contribution of parallel shunting resistivity in total resistivity can be expressed as:[69]

$$\frac{1}{\rho_{tot}(T)} = \frac{1}{\rho_i(T)} + \frac{1}{\rho_{shunt}} \qquad (7)$$

Where, $\rho_{tot}(T), \rho_i(T), and\ \rho_{shunt}$ represent the total, intrinsic, and shunting resistivity, respectively. The conventional electron-magnon scattering mechanism, typically observed in ferromagnets, fails to adequately explain the experimentally observed temperature dependence of resistivity in this case. In half-metallic ferromagnets, the existence of an energy gap at the Fermi level in one of the spin channels (commonly the minority spin channel) imposes a constraint on spin-flip scattering processes. Specifically, a minimum excitation energy of $K_B\Delta$ is required for majority spin electrons to transition into unoccupied minority spin states via spin-flip. Consequently, the standard quadratic electron-magnon scattering contribution to resistivity becomes exponentially suppressed at low temperatures. Thus, the conventional electron-magnon scattering term can be modified by incorporating a Boltzmann factor as:[60,71]

$$\rho_{e-m}(T) = \beta T^2 e^{-\Delta/T} \qquad (8)$$

where, $\Delta$ represents the energy gap between the Fermi level and the unoccupied band edge. An energy gap of $\Delta = 29.3$ meV and 30.85 meV were determined from the fitting for $x = 0.25$ and 0.375, respectively. The observed exponential suppression of single magnon scattering suggests the half metallic ferromagnetic nature of the samples.

The authors investigated the possibility of half-metallic ferromagnetism in Heusler alloys by analyzing low temperature resistivity (eliminating the shunting and electron-phonon terms) for two-magnon scattering by using the expression:[72–74]

$$\rho(T) = \rho_0 + \beta T^{9/2} \qquad (9)$$

The resistivity data of both samples fitted very well by $T^{9/2}$ term which suggests the strong contribution from two-magnon scattering as expected in half metallic materials. The author Aquil Ahmad *et al.* reported the two magnon scattering in full Heusler alloy $Co_2FeAl$ (nanoalloy).[75]

### 3.2 Magnetic and spin gapless semiconductors

Magnetic semiconductors (MSs)[76–80] and spin gapless semiconductors [81–84] (SGSs) can play an important role in spintronic devices as these materials have both magnetic and semiconducting properties and therefore acts as a bridge between magnetism and semiconductors. Such materials are expected to create new possibilities for spin-polarized charge transport and spin filtering via magnetic ordering and semiconducting characteristics. Understanding of their magnetic and transport properties is crucial step to design next-generation spintronic devices, such as spin injectors, magnetic tunnel junctions, and spin transistors.[76–78,85–87] MSs have non-zero but unequal gaps in both spin channels (spin up and down). On the other hand, SGSs have a gap in one spin channel, while another spin channel is gapless (the valence and conduction bands touch at the Fermi level). Schematics of DOS for SGSs and MSs are shown in Fig. 6 (b) and (c), respectively. Several magnetic semiconductors have been reported experimentally though reports on SGSs are only few. In the following sub-section, we would like to shine light on MSs and SGSs through some materials reported recently.

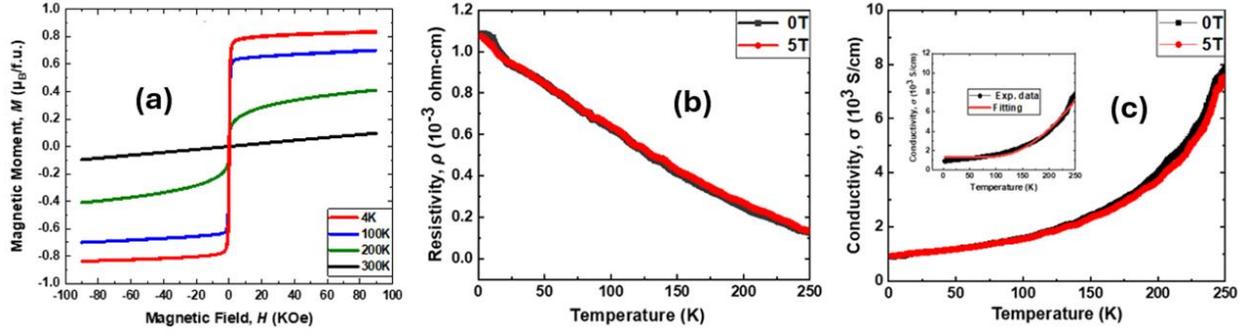

**Fig. 9.** (a) The field dependence of magnetization as a function of temperature for CoRuTiSn, (b) Electrical resistivity of CoRuTiSn with varying temperature at different fields, (c) the temperature dependence of electrical conductivity at different fields. Inset in (c) shows fitting of conductivity data by using equation (10). Reproduced with permission from R. Kumar *et al.* J. Alloys Compd. 1026 (2025) 180530 doi.org/10.1016/j.jallcom.2025.180530 © 2025 Elsevier.[88]

### 3.2.1 CoRuTiSn

Kumar *et al.* studied structural, magnetic and transport properties of quaternary Heusler alloy CoRuTiSn and reported the magnetic semiconducting behavior in the material.[88] CoRuTiSn was synthesized by arc-melting technique. The structural analysis reveals that the material crystallizes in tetragonal structure with small amounts of impurity. The magnetic field dependence of magnetization, shown in the Fig. 9(a) suggests the soft ferromagnetic nature of the material with $T_c$ ~200 K.[88] Magnetic moment of CoRuTiSn was determined to be 0.84 $\mu_B/f.u.$ at 4 K, which is slightly less than estimated by Slater-Pauling rule (1 $\mu_B$).[88] This deviation might be attributed to a tetragonal distortion/atomic disordering in the crystal structure of the material.

Fig. 9(b) and 9(c) depict the temperature dependence of electrical resistivity and electrical conductivity, respectively, with and without magnetic field. It is clear that the resistivity does not show any notable change with the magnetic field. The temperature dependence of resistivity shows decrease with increasing temperature, indicating semiconducting behavior. It can be carefully observed that resistivity shows exponential temperature dependence, ruling out the possibility of spin gapless properties, which generally observed to show linear temperature dependence. The temperature dependence conductivity $\sigma(T)$ data was fitted by the following equation and is shown in the inset of Fig. 9(c).[89,90]

$$\sigma(T) = \sigma_0 + \sigma_a e^{-E_g/K_B T} \qquad (10)$$

Where, $\sigma_0$ and $\sigma_a$ are used for residual conductivity due to non-stoichiometry/defects at low temperatures and zero-temperature conductivity for the semiconductor, respectively. Notation, $K_B$ and $E_g$ represent the Boltzmann constant and bandgap, respectively. The bandgap estimated by fitting equation (10) to experimental electrical conductivity data was found to be ~ 60.4 meV. The relatively high conductivity along with the small bandgap indicates that CoRuTiSn behaves as a narrow bandgap magnetic semiconductor. It is worth noting that theoretical calculations carried out by Bahramian and Ahmadian on CoRuTiSn predicted a half-metallic nature for the material, however the experimental results reveal its magnetic semiconducting behavior.[91] The possible reason could be the distortion of cubic structure to tetragonal and modification of electronic structures by small impurity phase.

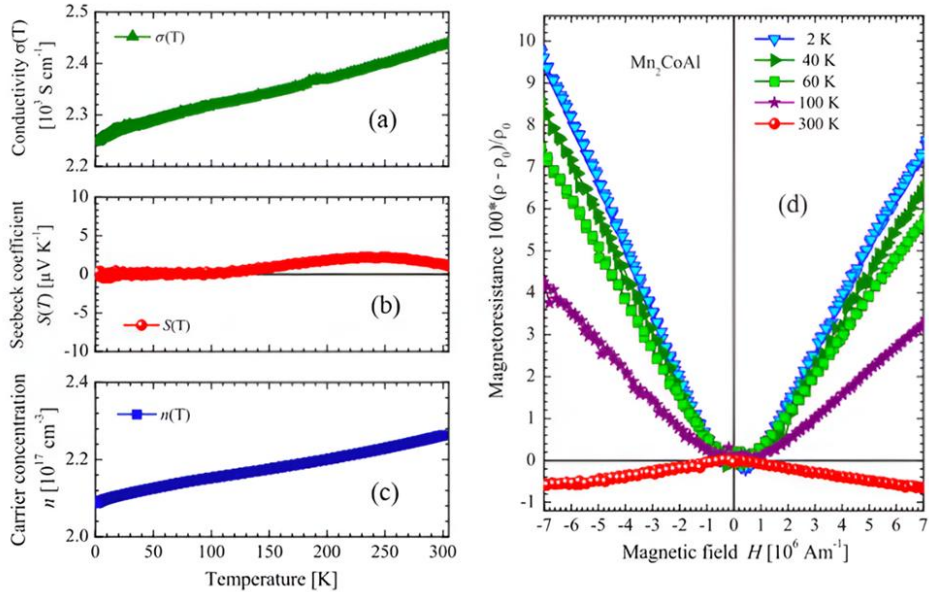

**Fig. 10.** (a) The temperature dependence of (a) electrical conductivity, (b) Seebeck coefficient, and (c) carrier concentration for Mn₂CoAl. (d) The field dependence of magnetoresistance as a function of temperature. Reproduced with permission from S. Ouardi *et al.* Phys. Rev. Lett. 110 (2013) 100401, doi.org/10.1103/PhysRevLett.110.100401. © 2013 American Physical Society.[93]

### 3.2.2 Mn₂CoAl

The electronic and magnetic properties of Mn₂CoAl were extensively studied by Liu *et. al.* through first-principles calculations.[92] Subsequently, Ouardi *et al.* experimentally validated the spin gapless semiconductor characteristics in polycrystalline bulk $Mn_2CoAl$, marking it the first experimental confirmation of SGS behavior in Heusler alloys.[93] Their investigations revealed that $Mn_2CoAl$ crystallizes in the inverse Heusler structure and exhibits a ferromagnetic behavior with saturation magnetization of 1.93 $\mu_B$/f.u. and a Curie temperature of 720 K. The electrical conductivity of $Mn_2CoAl$ demonstrated a typical behavior, diverging significantly from conventional metals or semiconductors. As shown in Fig. 10 (a), the electrical conductivity of the material increases nearly linearly with temperature, indicating non-metallic behavior. The temperature coefficient of resistivity (*TCR*) is notably low, measured to be $-1.4 \times 10^{-9}$ $\Omega \cdot$m/K. Furthermore, the conductivity remains nearly temperature-independent up to 300 K, attaining a value of 2440 S/cm. The Seebeck coefficient is negligible across a broad temperature range (5 K < T < 150 K), which has been attributed to electron-hole compensation as shown in Fig. 10 (b). At 300 K, the Seebeck coefficient remains exceptionally low, ~ 2 μV/K, a value inconsistent with typical semiconducting behavior. Fig. 10(c) further supports the gapless nature of the system by demonstrating a temperature-independent carrier concentration. Additionally, the magnetoresistance (*MR*) behavior of $Mn_2CoAl$ shows an unusual temperature-dependent sign reversal around 150 K as illustrated in Fig. 10(d). At low temperatures, the MR was non-saturating and nearly linear under high magnetic fields, resembling the behavior observed in gapless semiconductors known for their linear MR. At 2 K, a positive MR of ~10% is observed, whereas above 200 K, the MR becomes negative with low magnitude and exhibits a saturating trend with increasing magnetic field. The concurrence of temperature-independent conductivity and carrier concentration, negligible Seebeck coefficient, and linear *MR* at low temperatures are hallmark features of spin gapless semiconductor materials and confirm that $Mn_2CoAl$ is a SGS.[93]

### 3.2.3  CoFeMnSi

Following the experimental confirmation of the first Heusler spin gapless semiconductor, researchers began investigating other Heusler materials. Several experimental and theoretical studies have previously suggested that CoFeMnSi exhibits half-metallic behavior.[94] Xu *et al.* predicted the spin-gapless semiconductor nature of CoFeMnSi using first-principles calculations, identifying a half-metallic gap in one spin channel and a zero gap in the other.[81] Subsequently, Bainsla *et al.* confirmed the SGS behavior of CoFeMnSi through both experimental and theoretical investigations.[95]

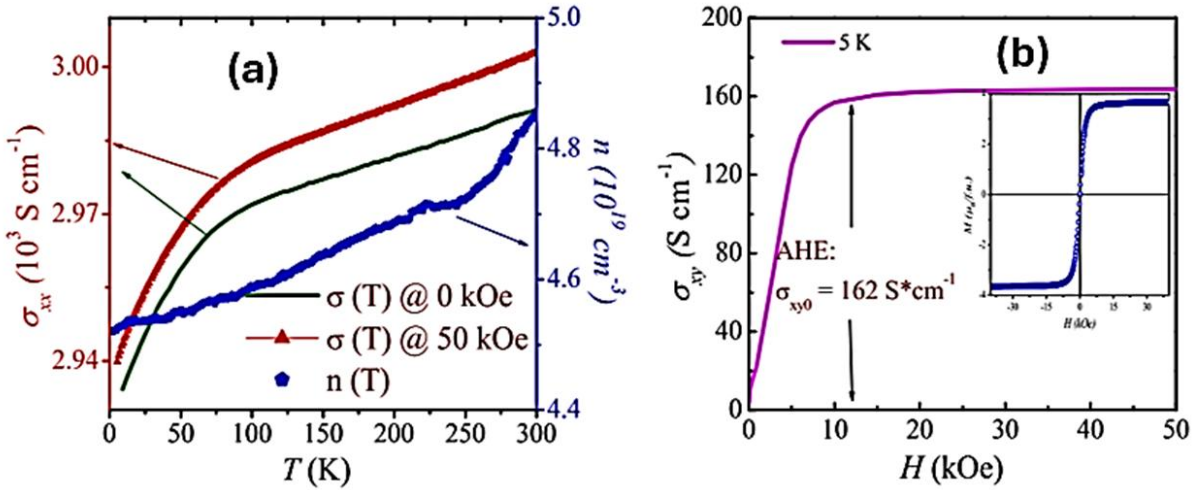

**Fig. 11.** (a) electrical conductivity and charge carrier concentration with varying temperature, (b) Hall conductivity vs. magnetic field at 5 K. Inset of (b) shows saturation magnetization curve at 5 K. Reproduced with permission from L. Bainsla *et al.*, Phys. Rev. B **91** (2015) 104408, doi.org/10.1103/PhysRevB.91.104408. © 2015 American Physical Society.[95]

The authors reported that the alloy crystallizes in a *DO₃*-type structure. The magnetic measurements show ferromagnetic behavior with saturation magnetization of $3.7\mu_B/f.u.$ and the Curie temperature of 620 K. Fig. 11(a) shows the temperature-dependent behavior of electrical conductivity and carrier concentration for CoFeMnSi. The conductivity increases with temperature, reflecting non-metallic behavior typical of spin gapless semiconductors. In the high-temperature regime, $\sigma(T)$ demonstrates a nearly linear temperature dependence. However, at low temperatures, the conductivity deviates from linearity, which is likely due to enhanced coherent scattering of conduction electrons caused by structural disorder. At 300 K, the conductivity was measured to be 2980 S/cm. The carrier concentration remains nearly constant over the temperature range of 5–300 K, strongly supporting the SGS behavior of CoFeMnSi. Fig. 11 (b) displays the Hall conductivity ($\sigma_{xy}$) as a function of the magnetic field at 5 K, with an anomalous Hall conductivity of 162 S/cm. All these results indicate the SGS nature of the material. In addition to the materials described in this section, several other materials have also been reported to exhibit SGS nature.[83,84,96–99]

### 3.3 Spin semi-metals

Semimetals bridge the gap between metals and semiconductors due to their unique electronic band structure. These materials possess intriguing features, such as low carrier density and high mobility, resulting from a slight overlap between the valence and conduction bands. Semimetals have been studied since the mid-20th century, with early research focusing on their transport and thermomagnetic properties (Delves, 1965).[100] Bismuth and graphite are classic examples that have been instrumental in understanding the behavior of materials with overlapping valence and conduction bands. Over the years, advancements in band structure calculations and experimental techniques have led to the discovery of novel classes of semimetals, including Dirac, Weyl, nodal-line semimetals, topological semimetals, and spin semimetals.[101–106] The exploration of these new classes of semimetals has stimulated extensive research into their potential applications in next-generation electronic and spintronic devices. In the following sub-section, we will discuss some newly discovered spin semimetal Heusler materials.

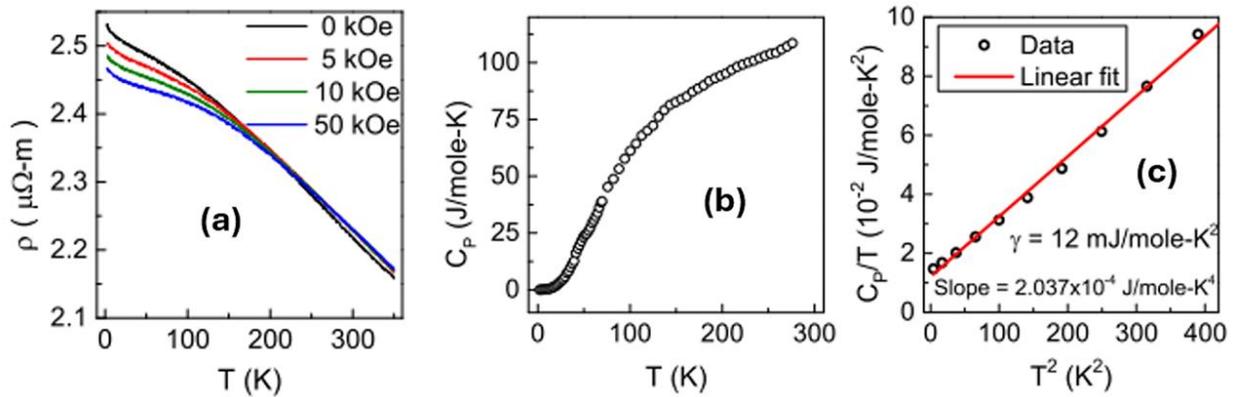

**Fig. 12.** (a) Electrical resistivity of FeRhCrGe with varying temperature at different fields, (b) the temperature dependence of specific heat, and (c) $C_p/T$ vs. $T^2$ with linear fitting. Reproduced with permission from Y. Venkateswara *et al.*, Phys. Rev. B 100 (2019) 180404(R), doi.org/10.1103/PhysRevB.100.180404. © 2019 American Physical Society.[107]

### 3.3.1 FeRhCrGe

Spin semimetals (SSMs) possess a unique band structure in which one spin channel (e.g., the majority spin) exhibits semimetallic behavior, while the other displays insulating or semiconducting characteristics (see Fig. 6(d)). Venkateswara *et al.* discovered and confirmed spin semimetal (SSM) behavior in the quaternary Heusler alloy FeRhCrGe through both theoretical and experimental studies.[107] This material crystalline in the $L2_1$ structure, which has 50% antisite disorder between tetrahedral site atoms, Fe and Rh. This material shows ferromagnetism with magnetic moment 2.9 $\mu_B/f.u.$ at 3 K. The Curie temperature of the material was found to be ~550 K.[107] Fig. 12(a) presents the temperature dependence of electrical resistivity under various magnetic fields for FeRhCrGe. The observed negative temperature coefficient of resistivity suggests possible semiconducting or semimetallic behavior. However, the absence of an exponential temperature dependence rules out a conventional gapped semiconductor. Consequently, the material is likely to be either a spin gapless semiconductor or a semimetal. The authors also performed specific heat ($C_p$) measurements, shown in Fig. 12(b) and (c). In conventional gapless semiconductors or spin gapless semiconductors, the electronic contribution to the specific heat is relatively insignificant compared to that in metals or semimetals, owing to the negligible density of states at the Fermi level ($E_F$). The authors did linear fitting of $C_p/T$ vs $T^2$ data which is illustrated in Fig. 12(c). Within the framework of the free electron gas model under the simplest approximation, the density of states at the Fermi level can be estimated as:[108]

$$n\,(E_F) = \frac{3\gamma}{\pi^2 K_B^2} \qquad\qquad (11)$$

here, γ represents the Sommerfeld constant. From the linear fitting, value obtained for γ was found to be 12 mJ/mole $K^2$ and the value obtained for $n\,(E_F)$ was found to be ~ 5.05 states/eV. A good correlation was observed between the experimental results and the theoretical calculations. Therefore, it is evident that FeRhCrGe does not exhibit the characteristics of a conventional gapless semiconductor or a spin gapless semiconductor; rather, it is most likely a spin semimetal.

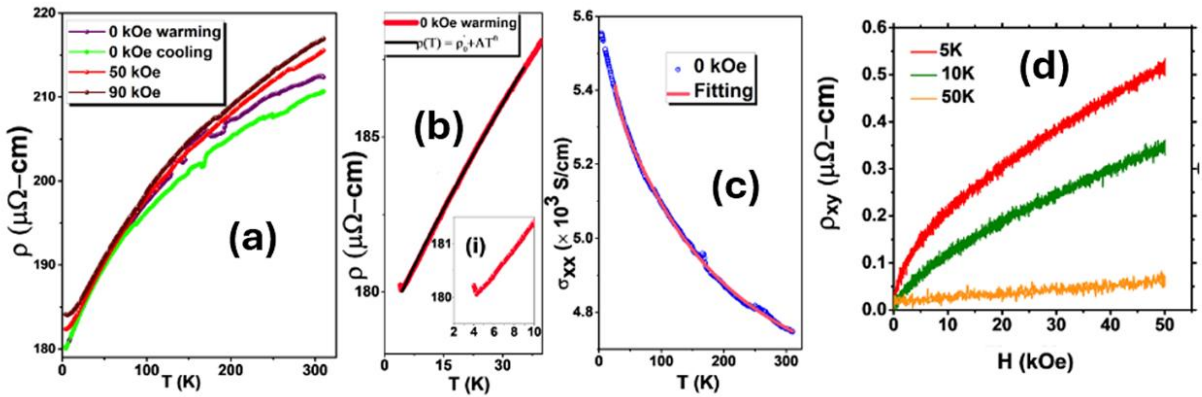

**Fig. 13.** (a) Temperature dependence of resistivity at various fields (b) fitting of resistivity, (c) fitting of conductivity (d) Hall resistivity vs. applied field. Reproduced with permission from J. Nag *et al*., ACS Appl. Electron. Mater. 5(2023) 5944. doi.org/10.1021/acsaelm.3c00935. © 2023 American Chemical Society.[102]

### 3.3.2   CoRuVSi

Nag *et. al.* reported detailed theoretical calculations and experimental results of spin semi-metal in quaternary Heusler alloy CoRuVSi.[102] The material crystallizes in the $L2_1$ structure with disorder between Co-Ru tetrahedral site atoms. According to the Slater–Pauling rule, CoRuVSi is expected to exhibit a saturation magnetization of 2.0 $\mu_B$ per formula unit in the fully ordered state. However, magnetization measurements at 3 K reveal a substantially reduced saturation magnetization of only 0.13 $\mu_B/f.u.$, representing a pronounced deviation from the Slater–Pauling prediction.[102] The temperature dependence of resistivity for CoRuVSi is shown in Fig. 13(a), which reveals semi-metallic behavior, as further corroborated by electronic structure calculations. Resistivity data was fitted with power law equation (4) in the temperature range of 5-30 K, shown in Fig 13(b). To gain deeper insights, the electrical conductivity data was analyzed using a modified two-carrier model within the temperature range of 30–310 K, shown in Fig. 13 (c). The two-carrier model for $\sigma(T)$ is described as:[109]

$$\sigma(T) = e\,(n_e\,\mu_e + n_h\,\mu_h) \tag{11}$$

here, $n_i = n_{i0}\,exp\,\frac{-\Delta E_i}{K_B T}$, $(i = e,\,h)$ are the electron/hole carrier concentrations with mobilities $\mu_i$ and pseudo- energy gaps $\Delta E_i$. The equation can be re-written as:

$$\sigma(T) = \left[A_e(T)\,exp\,\frac{-\Delta E_e}{K_B T} + A_h(T)\,exp\,\frac{-\Delta E_h}{K_B T}\right] \tag{12}$$

The pseudo-energy gaps for electrons and holes were determined to be 0.11 and 15.9 meV, respectively, indicating their small magnitude than narrow band gap semiconductors. This suggests that atomic disorder plays a vital role in substantially reducing the pseudo-energy gaps, particularly for electrons, which exhibit an exceptionally small gap. Such a narrow gap renders the electronic states highly susceptible to transitioning toward metallic behavior under minor perturbations, such as applied external fields or thermal fluctuations.

The field-dependence of Hall resistivity at different $T$ is displayed in Fig. 13 (d). Total Hall resistivity of magnetic materials can be expressed as:[17]

$$\rho_{xy}\,(T) = \rho_{xy}^{O} + \rho_{xy}^{A} = R_0 H + R_A M_S \qquad (13)$$

notation $\rho_{xy}^{O}$ and $\rho_{xy}^{A}$ represent the ordinary and anomalous contribution, $R_0$, $R_A$ and $M_S$ denote the ordinary, anomalous Hall coefficients and saturation magnetization of material, respectively. Anomalous Hall contribution observed only for 5 and 10 K, on the other hand Hall conductivity drop nearly to zero at 50 K, demonstrates the anomalous contribution disappeared at higher temperatures. Contribution of anomalous Hall resistivity was found to be 0.15μΩ-cm. Field-dependence of anomalous Hall conductivity at 5K was found 45 S/cm. At 5 K, the measured carrier concentration ($n$) is $7.4 \times 10^{18}$ cm$^{-3}$, which is well within the range of semimetal/semiconductor carrier densities. The authors also performed theoretical calculation and used other experimental tools to confirm the spin semimetal behavior of the CoRuVSi.[102]

### 3.4 Nearly zero moment materials

Fully compensated ferrimagnets (FCFs), also known as half-metallic antiferromagnets, represent an exciting class of materials in which the net magnetic moment is canceled due to perfect compensation between the opposing magnetic moments of the sublattices.[110–113] Leuken and de Groot theoretically demonstrated that this novel class of materials could exhibit 100% spin polarization without possessing a net magnetic moment, coining the term "half-metallic antiferromagnets (AFMs)" for such systems.[114] This classification was later refined and renamed "half-metallic fully compensated ferrimagnets (HM-FCFs). Although both antiferromagnetic materials and fully compensated ferrimagnets exhibit zero or nearly zero net magnetic moment, the origin of this behavior is fundamentally different. In antiferromagnets, the spin-up and spin-down band densities of states are symmetric, resulting in equal contributions from both spin channels to electrical conductivity. This symmetry leads to a zero net spin-polarized current. In contrast, FCF materials exhibit ferromagnetic ordering at the microscopic level, with magnetic moments aligned in such a way that they cancel each other out, producing no net magnetization. FCFs typically require three or more magnetic ions to break inversion symmetry, whereas AFM materials usually contain an even number of magnetic ions to maintain it. For spintronic applications involving ferromagnetic, ferrimagnetic, or fully compensated ferrimagnetic materials, breaking inversion symmetry in the magnetic structure is essential.[110] Antiferromagnets, which require magnetic inversion symmetry, should not be confused with fully compensated ferrimagnets, a specific type of ferrimagnet.[40] In the following, we will discuss some novel materials exhibiting the FCF behavior.

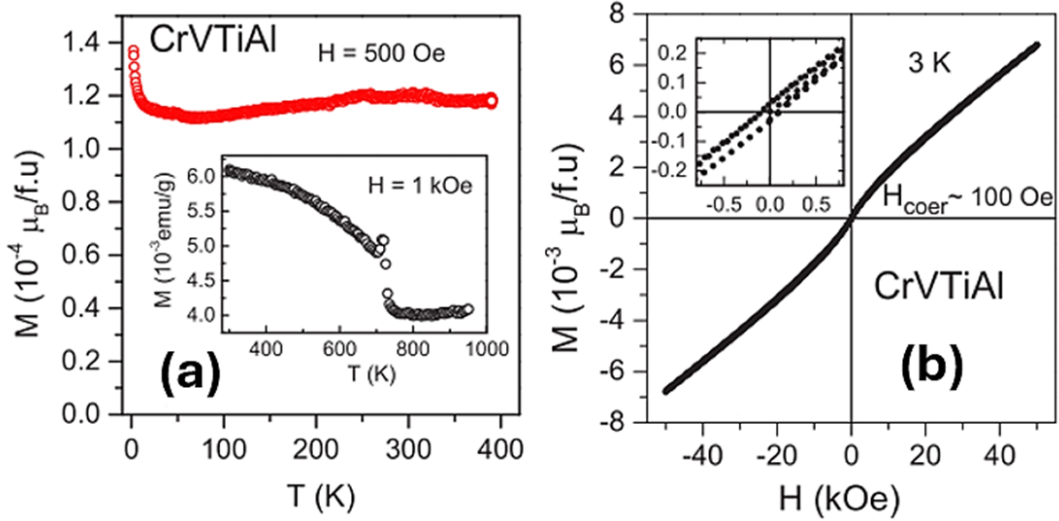

**Fig. 14.** (a) The temperature dependence magnetization (*M-T*) data up to 400 K, inset of (a) shows the *M-T* from 400-800 K, (b) the isothermal magnetization data at 3 K, inset of (b) shows the M-H on zoomed scale. Reproduced with permission from Y. Venkateswara *et al*., Phys. Rev. B 97 (2018) 054407, doi.org/10.1103/PhysRevB.97.054407. © 2018 American Physical Society.[40]

### 3.4.1 CrVTiAl

Theoretical calculations performed by some researchers suggested CrVTiAl to be a FCF[115,116] which was experimentally confirmed by Venkateswara *et al*.[40] CrVTiAl was synthesized by arc melt technique, crystallizes in $Y_{II}$ configuration with antisite disorder of Al with Cr and V. CrVTiAl has 18 valance electrons and as per *S-P* rule ($M = N_V$~18 rule, where $N_V$ is no. of valence electrons) magnetic moment should be zero. Fig. 14(a) and (b) shows the thermomagnetic curve at 1kOe field and isothermal magnetization at 3K by applying 100Oe field, respectively. A very small moment of CrVTiAl (about $10^{-3}\mu_B/f.u.$) and a high magnetic ordering temperature (710 K) were observed from these results. The authors reported that the observed non-zero hysteresis may arise from the distinct temperature dependencies of Cr, V, and Ti sublattices. Another possibility may be due to small deviations from the ideal 1:1:1:1 stoichiometry. On the basis of experimental and theoretical investigations, the author confirmed the CrVTiAl to be FCF material. Several other Heusler alloys were also reported to show FCF nature and could be promising for antiferromagnetic spintronics.[110,111,113,117,118]

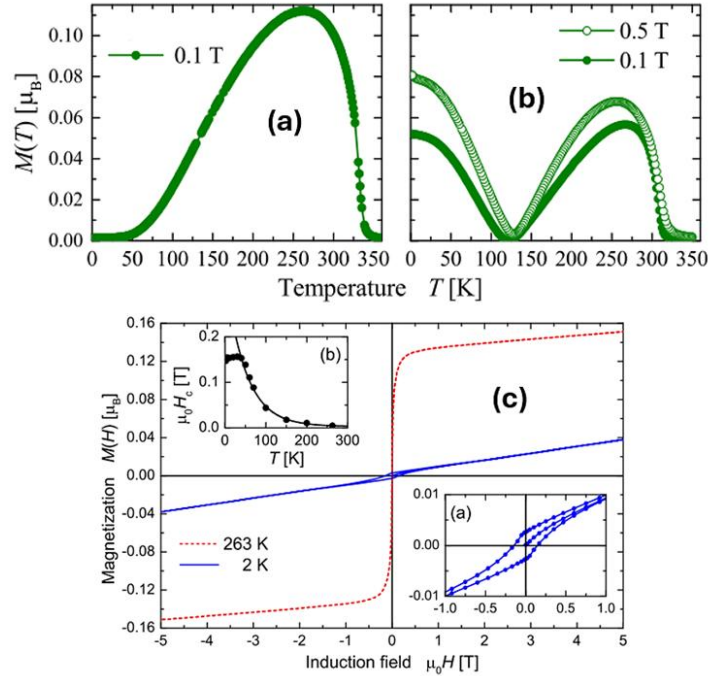

**Fig. 15.** (a) The magnetization curve with varying temperature at 0.1 T field for fully compensated ferrimagnet, (b) the magnetization with varying temperature at 0.1 and 0.5 T field for an overcompensated sample, (c) the magnetization data with varying field for $Mn_{1.5}FeV_{0.5}Al$ at 2 and 263 K, inset of in (c), (a) the zoomed scale of magnetization at low temperature, inset (b) temperature dependence of the coercive field. Reproduced with permission from R. Stinshoff *et al.*, Phys. Rev. B **95** (2017) 060410(R), doi.org/10.1103/PhysRevB.95.060410. © 2017 American Physical Society.[117]

### 3.4.2 $Mn_{1.5}FeV_{0.5}Al$

Stinshoff *et al.* reported fully compensated ferrimagnetic behavior in $Mn_{1.5}FeV_{0.5}Al$, synthesized by arc melt technique.[117] The material has $L2_1$ crystal structure. Figure 15(a) illustrates the temperature dependence of magnetization for a FCF. As expected for a fully compensated ferrimagnet, the magnetization approaches zero near 0 K and remains nearly zero up to approximately 50 K. A clear onset of magnetic ordering is observed at the Curie temperature, which appears around 335 K. In contrast, Fig. 15(b) presents $M(T)$ curves for a slightly overcompensated sample measured under various magnetic fields. From the data obtained at 0.1 T, a Curie temperature of approximately 308 K is determined.[117] Interestingly, as temperature decreases, the magnetization initially diminishes to zero at around 127 K and subsequently increases with further cooling. This characteristic behavior signifies the presence of a magnetic compensation point within the ferrimagnetic phase. Notably, the fully compensated magnetic state is extremely sensitive to the sample's compositional stoichiometry. Fig. 15(c) depicts the magnetization as a function of applied magnetic field at two temperatures; 263 K, where the magnetization reaches its maximum, and 2 K, where it is nearly zero. At higher temperatures, the material exhibits soft magnetic behavior, whereas at low temperatures, a small but finite remanence and coercive field are observed.[117] As shown in inset (b), the coercive field remains constant below 50 K, consistent with the vanishing magnetization in this regime [compared to Fig. 15(a)]. Above this threshold temperature, the material becomes magnetically softer with increasing temperature. The presence of a coercive field near the compensation point is a well-known phenomenon in ferrimagnetic systems.[119] In some cases, the coercive field is predicted to diverge at the compensation point; however, in the case of a fully compensated half-metallic ferrimagnet, it clearly saturates below the critical temperature.[117]

### 3.4.3 Cr$_2$MnSb

Several materials have been theoretically predicted to be fully compensated ferrimagnets; however, only a few have been experimentally realized, most of which are in bulk form. To integrate these materials into spintronic devices, it is crucial to grow them in thin-film form. Only a limited number of studies in the literature report on the thin-film growth and characterization of these materials. Here we discuss Cr$_2$MnSb thin films reported to show FCF behavior.

Gupta *et al.* deposited 30 nm thick Cr$_2$MnSb on MgO(001) substrate using magnetron sputtering.[35] As deposited samples were annealed at various temperatures to improve the crystallinity of the sample. It has been observed that below 300 ˚C, the Cr$_2$MnSb films show amorphous nature and crystallizes in full $B2$ and partial $L2_1$ ordering after the annealing above 300 °C. The authors performed magnetic and transport measurements on the samples annealed at and above 300 °C. Fig. 16(a) shows the in-plane field dependence of magnetization at different annealing temperatures, revealing that the magnetization increases with increasing temperatures.

The film annealed at 300 °C shows the lowest moment of 0.3 $\mu_B$/f.u. with a Curie temperature of ~350 K (Fig. 16 (a & b)). Fig. 16(b) shows magnetization at 10 K with magnetic field applied in in-plane and out of-plane directions. Magnetization under the in-plane and out of-plane magnetic fields are similar, suggesting that the film shows very weak magnetic anisotropy. Fig. 16 (d & e) shows the annealing temperature dependence of magnetic moment and coercive field at 10 K, measured under in-plane field of 1 T.

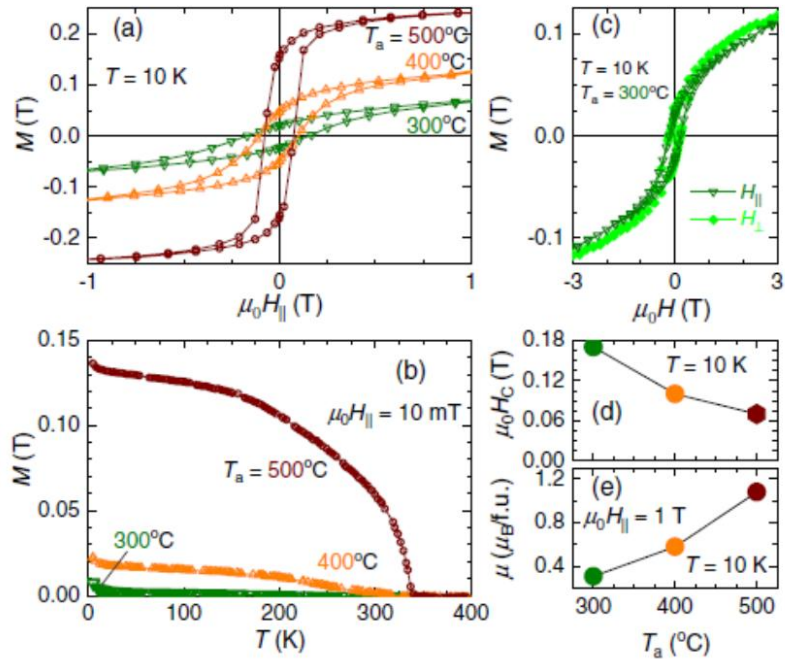

**Fig. 16.** (a) The magnetization data of $Cr_2MnSb$ at different annealing temperatures in in-plane magnetic field, (b) magnetization vs. temperature of $Cr_2MnSb$ for different annealing temperatures, (c) magnetization curves of $Cr_2MnSb$ for in-plane and out of plane magnetic fields at 10 K, (d) and (e) show the coercive field and magnetic moment at 1 T fields for different annealing temperatures, respectively. Reproduced with permission from S. Gupta *et al.*, J. Phys. D. Appl. Phys. 52 (2019) 495002, doi.org/10.1088/1361-6463/ab3fc6. © 2019 IOP Publishing.[35]

After confirmation of FCF behavior from the magnetization results, the authors integrated $Cr_2MnSb$ film as one of the electrodes for magnetic tunnel junction (MTJ) to check the potential of these materials in spintronic technology. For this the authors fabricated MTJ devices on $Si/SiO_2$ substrate, and measured junction resistance as a function of magnetic field. The results are shown in Fig. 17 for blanket film and MTJ devices. The MTJ devices show clear switching (Fig. 17 (c &

d)), however the magnitude of tunnel magnet- resistance (TMR) was small. The small TMR effect might be attributed to several reasons such as interface quality, crystallinity of the sample and spin polarization of the electrode materials.[35] The authors also performed cross-sectional transmission electron microscopy to investigate some of the reasons behind low TMR, which can be seen in Fig. 17 (e & f).

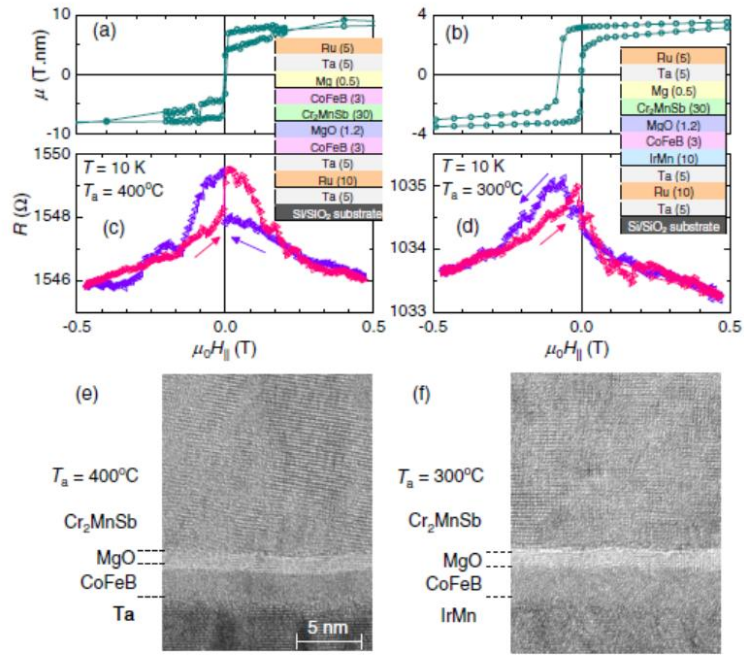

**Fig. 17.** Field dependence of magnetization for blanket films at (a) 400 °C and (b) 300 °C annealing temperatures. The field dependence of junction resistance for MTJ devices at (c) 400 °C and (d) 300 °C annealing temperatures. (e) and (f) show the samples image taken by transmission electron microscope. Reproduced with permission from S. Gupta *et al.*, J. Phys. D. Appl. Phys. 52 (2019) 495002, doi.org/10.1088/1361-6463/ab3fc6. © 2019 IOP Publishing.[35]

# 4. Applications

Half-metallic ferromagnets (HMFs), characterized by 100% spin polarization at the Fermi level, are highly promising for spintronic applications.[120–123] They serve as efficient spin injectors in spin valves, magnetic tunnel junctions (MTJs), and giant magnetoresistance (GMR) devices, and are integral to non-volatile magnetic random-access memory (MRAM).[35,124–127] In addition, HMFs enhance spin injection efficiency in spin filters and spin transistors, thanks to their robust magnetic ordering and high Curie temperatures.[5,128]

Magnetic semiconductors, on the other hand, are crucial for the development of spintronic semiconductor technology.[77,79] Their inherent magnetic nature, combined with semiconducting properties, makes them excellent candidates for spin filters, facilitating efficient spin injection into conventional semiconductors with minimal lattice mismatch.[76,77,85] These materials are also suitable for use in spin field-effect transistors (Spin-FETs), magneto-optical devices, and spin light-emitting diodes (Spin-LEDs), which emit circularly polarized light and offer exciting possibilities for opto-spintronic applications.[85] Whereas spin gapless semiconductors (SGSs) uniquely combine high charge carrier mobility with 100% spin polarization. The tunability of their electronic structure allows multifunctional operation, making them attractive for both logic and sensing technologies.[12,87,93,129,130]

Spin semimetals (SSMs), including magnetic Weyl and Dirac semimetals, exhibit exotic transport properties such as the anomalous Hall effect, high mobility, and chiral anomaly, all with minimal energy dissipation.[110–113] These materials are well suited for quantum spin Hall devices, topological spintronics, and next-generation logic and memory architectures that exploit the synergy between topological protection and magnetic order.[101,131–133] Fully compensated ferrimagnets (FCFs) stand out due to their zero net magnetization, which minimizes stray magnetic fields. This makes them ideal for high-density magnetic storage and low-noise spintronic applications.[110,111,113,117,118] Despite having microscopic ferromagnetic ordering, the complete cancellation of magnetic moments between sublattices ensures stable, low-power, and thermally robust device operation. FCFs also enable fast spin switching and are particularly promising for advanced MTJ-based MRAM and antiferromagnetic spintronic technologies.[35,112]

## 5. Summary and outlook

Heusler alloys have emerged as a highly promising class of materials for spintronic applications due to their tunable electronic and magnetic properties. Their ability to exhibit diverse functionalities—such as half-metallicity, spin gapless semiconducting behavior, and fully compensated ferrimagnetism—makes them ideal candidates for next-generation spintronic devices. These materials offer a versatile platform for developing energy-efficient, scalable, and high-speed devices, thereby paving the way for future information technologies. Despite their significant potential, several challenges must be addressed before these materials can be fully integrated into practical spintronic applications. One major hurdle is their sensitivity to disorder, as the electronic structure of Heusler alloys is highly tunable and can be easily altered by even minor impurity/disorder. Furthermore, while thousands of Heusler alloys have been predicted theoretically, only a limited number have been successfully synthesized and characterized experimentally. Most of these are in bulk form, which poses limitations for device fabrication. For successful integration into spintronic devices, it is crucial to develop and study these materials in thin-film form. Thin-film growth presents its own challenges but remains a key step toward practical implementation. Moreover, the novel properties such as half metallicity and spin gapless semiconducting of these materials is often inferred from indirect measurements such as electrical transport. To gain a more accurate understanding of their electronic structure, it is essential to employ direct experimental techniques like angle-resolved photoemission spectroscopy (ARPES). Going forward, continued efforts are required to discover and synthesize new Heusler compounds, especially in thin-film form. Comprehensive experimental characterization using both direct and indirect techniques will be critical in confirming their predicted properties and advancing their application in real-world spintronic devices.

## Acknowledgement

S.G. acknowledges the financial support from Anusandhan National Research Foundation (ANRF), New Delhi under the project SUR/2022/004713.